\documentclass[report,11pt]{article}

\usepackage{scrextend}
\usepackage{amsmath,amssymb,epsfig,bbm}
\usepackage{MnSymbol}
\usepackage{slashed}
\usepackage{amsthm}

\usepackage{float}

\usepackage{enumerate}

\usepackage{color}
\usepackage{mathrsfs}

\usepackage{dsfont}

\definecolor{co}{cmyk}{0,0.7,0.3,0}
\definecolor{darkgreen}{cmyk}{1,0,1,.2}
\definecolor{m}{rgb}{1,0.1,1}

\newcommand{\be}{\begin{equation}}
\newcommand{\ba}{\begin{eqnarray}}
\newcommand{\ea}{\end{eqnarray}}
\newcommand{\nn}{\nonumber}

\def\d{\delta}

\def\p{\pi}

\def\OO{\Omega}

\def\X{\Xi}

\def\ca{{\cal A}}

\def\cd{{\cal D}}

\def\cf{{\cal F}}
\def\cg{{\cal G}}
\def\ch{{\cal H}}

\def\cj{{\cal J}}

\def\cm{{\cal M}}

\makeatletter
\newcommand{\eqnum}{\refstepcounter{equation}\textup{\tagform@{\theequation}}}
\makeatother

\newcommand{\pa}{\partial}

\newtheorem*{definition*}{Definition}

\fontfamily{yfrak}

\begin{document}

\vskip 25mm

\begin{center}

{\large\bfseries

Dirac Operators on Configuration Spaces: Fermions with Half-integer Spin, Real Structure, and Yang-Mills Quantum Field Theory


}

\vskip 6ex

Johannes \textsc{Aastrup}$^{a}$\footnote{email: \texttt{aastrup@math.uni-hannover.de}} \&
Jesper M\o ller \textsc{Grimstrup}$^{b}$\footnote{email: \texttt{jesper.grimstrup@gmail.com}}\\ 
\vskip 3ex

$^{a}\,$\textit{Mathematisches Institut, Universit\"at Hannover, \\ Welfengarten 1, 
D-30167 Hannover, Germany.}
\\[3ex]
$^{b}\,$\textit{Copenhagen, Denmark.}
\\[3ex]

{\footnotesize\it This work is financially supported by entrepreneur Kasper Gevaldig, Denmark,\\ and by Master of Science in Engineering Vladimir Zakharov, Granada, Spain.}

\end{center}

\vskip 3ex

\begin{abstract}

\vspace{0.5cm}

In this paper we continue the development of a spectral triple-like construction on a configuration space of gauge connections. We have previously shown that key elements of bosonic and fermionic quantum field theory emerge from such a geometrical framework. In this paper we solve a central problem concerning the inclusion of fermions with half-integer spin into this framework. We map the tangent space of the configuration space into a similar space based on spinors and use this map to construct a Dirac operator on the configuration space. We also construct a real structure acting in a Hilbert space over the configuration space.
Finally, we show that the self-dual and anti-self-dual sectors of the Hamiltonian of a non-perturbative quantum Yang-Mills theory emerge from a unitary transformation of a Dirac equation on a configuration space of gauge fields. The dual and anti-dual sectors emerge in a two-by-two matrix structure.

\end{abstract}

\newpage


\section{Introduction}

One of the key tasks in the search for a fundamental theory is to identify hypotheses that on the one hand are based on extremely simple principles -- this enhances their immunity to further scientific reductions and hence their chances of being fundamental -- and on the other hand are capable of generating rich mathematical structures that match the mathematics that we encounter in modern high-energy physics.

During the past two decades we have developed such a hypothesis \cite{Aastrup:2005yk}-\cite{Aastrup:2023jfw}. Our proposal is to start with an algebra of holonomy-diffeomorphisms -- this is the so-called $\mathbf{HD}$-algebra \cite{Aastrup:2012vq,AGnew}, which is generated by parallel transports along flows of vector-fields on a three-dimensional manifold -- and to consider the geometry of the corresponding configuration space of spin-connections. Since the $\mathbf{HD}$-algebra comes with a very high level of canonicity -- essentially it only depends on the dimension of space -- this hypothesis has a high level of irreducibility in terms of further scientific reductions.

Concretely, the idea is to construct either a Dirac or a Bott-Dirac operator on the configuration space and to employ the machinery of noncommutative geometry  \cite{Connes:1996gi}-\cite{1414300} to the combined system of Dirac or Bott-Dirac operator and the noncommutative $\mathbf{HD}$-algebra. We have previously shown that the basic building blocks of bosonic and fermionic quantum field theory emerges from such a geometrical framework: 
\begin{enumerate}
\item[-]
the Hamiltonian operators of a Yang-Mills quantum field theory coupled to a fermionic sector emerges from the square of a Bott-Dirac operator defined on a configuration space \cite{Aastrup:2020jcf,Aastrup:2023jfw},
\item[-]
the interaction between the Dirac operator and the $\mathbf{HD}$-algebra encodes the canonical commutation relations of a quantum gauge theory \cite{Aastrup:2020jcf,Aastrup:2014ppa},
\item[-]
the canonical anti-commutation relations of a quantised fermionic field emerges from the CAR algebra that we used to construct the Dirac operator \cite{Aastrup:2020jcf},
\item[-]
while all of the above is formulated on a curved, dynamical background \cite{Aastrup:2020jcf,Aastrup:2023jfw}. 
\end{enumerate}

One of the missing pieces in the development of this framework is the inclusion of half-integer fermions. Since the configuration space involves spin-one objects it is not obvious how a derivation on this space can be coupled to an infinite-dimensional Clifford algebra based on half-integer fields in a natural way. In this paper we solve this problem by mapping the tangent space on the configuration space into a similar space based on spinors.
Furthermore, once we have half-integer spin fermions it is natural to introduce a real structure, which we, in turn, use to define the Dirac operator on the configuration space.

We then show that the selfdual and anti-selfdual sectors of a Yang-Mills quantum field theory emerges from a Dirac operator on the configuration space via a unitary transformation that involves the Chern-Simons term. 
%
%
%
Concretely, we find that the self-dual and anti-self-dual sectors emerge in a two-by-two matrix structure obtained from the square of a unitarily rotated Dirac operator. Alongside the Hamiltonian we also find a spectral invariant, that measures the assymmetry of the spectrum of a covariant derivative on the underlying manifold. 

All this shows that there is a direct connection between nonperturbative quantum gauge theory and noncommutative geometry of configuration spaces, a realisation that calls for a deeper understanding of the geometry of these spaces. 
This reaches beyond the scope of quantum gauge theory itself: As soon as a Dirac operator that interacts with a noncommutative algebra (such as the $\mathbf{HD}$-algebra) has been introduced one is in the domain of noncommutative geometry, which in its core is a framework of unification.

The relevance of noncommutative geometry to high-energy physics is due to Chamseddine and Connes' work on the standard model of particle physics. They have shown that the standard model coupled to general relativity can be formulated as a certain almost-commutative spectral triple on the underlying four-dimensional manifold \cite{Connes:1996gi},\cite{Chamseddine:1996zu}-\cite{Barrett:2006qq} (see also \cite{Dubois-Violette:1988hil}). This work, which casts the standard model in a completely new conceptual light, comes, however, with a number of challenges, where arguable the most important one is how to include quantum field theory in a natural way. The point is that since Chamseddine and Connes' work is inherently gravitational it {\it cannot} be quantised in its entirety in any conventional way since this would include a quantisation of gravity.  
One of the initial motivations behind our reseach project \cite{Aastrup:2005yk} was precisely to address this problem of how to incorporate non-perturbative quantum field theory into a framework based on noncommutative geometry. 

The notion of a geometry of configuration spaces of gauge connections is, however, not new but was considered already by Feynman \cite{Feynman:1981ss} and Singer \cite{Singer:1981xw} (see also \cite{Orland:1996hm}), but the idea to study non-trivial geometries and in particular to study their dynamics, is new.  

 This paper is organised as follows: 
In section 2 we set the stage for a geometrical construction on a configuration space of gauge connections and show how an infinite-dimensional Clifford algebra based on half-integer spin objects is constructed. In section 3 we then introduce a Dirac operator on the configuration space and in section 4 we show that the square of a Dirac operator, which is obtained through a unitary transformation of the original Dirac operator, leads to a Hamilton operator of a Yang-Mills quantum theory. We end the paper in section 5 with a discussion.

\section{Metrics on a configuration spaces and fermions with half-integer spin}

Let $M$ be a three-dimensional manifold. We will for simplicity assume that $TM$ is trivializable. We will also assume that we have a Riemannian metric on $M$. This induces a metric on $TM$. We will later discuss how much the subsequent construction depends on this Riemannian metric. The Riemannian metric in turn gives rise to a Clifford bundle $Cl (TM)$.  Note that fiberwise $Cl (TM)$ is isomorphic to $\mathbb{H}\oplus \mathbb{H}$, and if we complexify, i.e. consider  $\mathbb{C}l (TM)=Cl (TM)\otimes_{\mathbb{R}}\mathbb{C}$, then this becomes fiberwise isomorphic to $M_2(\mathbb{C})\oplus M_2(\mathbb{C})$. Thus we have fiberwise two irreducible copies of $Cl (TM)$.  We thus get two spin$^\mathbb{C}$-bundles, $S_1$ and $S_2$, both fiberwise isomorphic to $\mathbb{C}^2$.   

Let $\ca$ be the space of smooth spin connections.  Since spin$(2)=SU(2)$ the space $\ca$ consists of $\mathfrak{su}(2)$-connections.   We can choose a trivialization of $S_1$ and $S_2$ such that the elements in $\ca$ acts on each of $S_1$ and $S_2$ as $\mathfrak{su}(2)$-connections. 

On $\mathbb{C}l (TM)$ we have a real structure in the following way: For en element $x\otimes \lambda \in \mathbb{C}l (TM)=Cl (TM)\otimes_{\mathbb{R}}\mathbb{C}$ we define 
$$ \overline{x\otimes \lambda}=x\otimes \overline{\lambda} .$$ This real structure induces a  charge conjugation operator (see \cite{Meyer:2001}) 
$$ C:S_1\oplus S_2 \to S_1\oplus S_2    ,$$
which is a fiberwise antilinear isometry acting diagonally with 
$$
C (x\otimes \lambda )C^*=\overline{x\otimes \lambda} 
$$ 
and $C^2=-1$. 
Furthermore $C$ commutes with the action of $Cl(TM)$.

We can also enrich $TM$ with an extra orthogonal direction, in this way we get bundles $Cl^4(TM)$ and $\mathbb{C}l^4(TM)$, with fibers $Cl(4)$ and $\mathbb{C}l(4)$. Since $\mathbb{C}l(4) =M_n(\mathbb{C})$ the representation of  $\mathbb{C}l(TM)$ on $S_1\oplus S_2$   extends to a representation of $\mathbb{C}l^4(TM)$ on $S_1\oplus S_2$. Also in this case $C$ commutes with $Cl^4(TM)$.

\subsection{A Hilbert space over $\ca$}

In order to onstruct a Dirac operator over $\ca$ we need a Hilbert space $L^2(\ca)$. In this paper we shall not concern ourselves with the details on how this Hilbert space is constructed but simply refer the reader to \cite{Aastrup:2020jcf} and \cite{Aastrup:2023jfw}. The only point that we need to mention is that the construction of $L^2(\ca)$ requires a choice of gauge fixing $\cf$ on $\ca$, which means that we require that for each $\nabla \in \ca$ there is exactly one $g\in \cg$ with $g(\nabla) \in \cf $, $\cg$ being the space of gauge transformations. The construction of the Hilbert space $L^2(\cf)$ then involves a BRST quantisation procedure. 

In the following we shall work only with $\cf$ instead of $\ca$ and ignore all issues that might emerge from this gauge fixing. This includes in particular any issues related to the tangent space over $\ca$ and with constructing derivates on $\ca$. Again, we refer the reader to \cite{Aastrup:2023jfw} for details.

\subsection{Mapping into one-forms with values in the spin bundle}

In order to formulate the tangent space over $\cf$ note that if we choose a connection  $ \nabla_0 \in  \cf$, then we can write any connection in $\cf$ on the form $\nabla = \nabla_0 + \omega$, where $\omega \in \Omega^1 (M,\mathfrak{g})$. In this way we can write the tangent space as \cite{Aastrup:2020jcf}
$$
T\cf =\cf \times \Omega^1 (M,\mathfrak{g}) .
$$   
We wish to interpret elements in the tangent space  $T_{\nabla}\cf= \Omega^1 (M,\mathfrak{g})$ as fermions. This interpretation is hindered, however, by the fact that elements in $\Omega^1 (M,\mathfrak{g})$ have integer spin. In order to construct elements with half integer spin we are going to construct a map 
$$\cf \times \Omega^1 (M ,\mathfrak{g} ) \to \cf \times \Omega^1 (M ,S_1\oplus S_2 ) $$
fibered over $\cf$. We first note that there is a map 
$$ P:\Omega^1 (M ,\mathfrak{g} )  \times C^\infty (M,S_1\oplus S_2) \to  \Omega^1 (M ,S_1\oplus S_2 ),$$
since $\mathfrak{g}$ is acting on $S_1$ und $S_2$. Next we choose $(\psi_1,\psi_2) \in C^\infty(M,S_1\oplus S_2)$. With the above map we get a map 
$$ \chi_{(\psi_1,\psi_2) }:\Omega^1 (M,\mathfrak{g})\to \Omega^1 (M,S_1\oplus S_2)$$
given by 
$$ \chi_{(\psi_1,\psi_2) } (\omega) =P(\omega,(\psi_1,\psi_2)  ).$$
Finally the map 
$$ \Psi_{(\psi_1,\psi_2) }  :\cf \times \Omega^1 (M ,\mathfrak{g} ) \to \cf \times \Omega^1 (M ,S_1\oplus S_2 ) $$
is just given by 
$$ \Psi_{(\psi_1,\psi_2) } (\nabla, \omega)=(\nabla , \chi_{(\psi_1,\psi_2) } (\omega) ). $$

\subsection{Metrics on $\cf \times \Omega^1 (M ,\mathfrak{g} ) $}

Next we want to discuss metrics on $\cf \times \Omega^1 (M ,S_1\oplus S_2 )$. 
We want to construct them as metrics fibered over $\cf$, i.e. we want metrics of the type 
\begin{equation}
\cf\ni \nabla \to \langle \cdot , \cdot \rangle_{\nabla} ,
\label{innerspinor}
\end{equation}
where $\langle \cdot , \cdot \rangle_{\nabla}$ is an inner product on $\Omega^1(M,S_1\oplus S_2 )$.

The first remark is that if we have a metric on 
$\cf \times \Omega^1 (M ,S_1\oplus S_2 )$, which is complex,  we can pull it back to a metric on  $\cf \times \Omega^1 (M ,\mathfrak{g} )$ with the map  $\Psi_{(\psi_1,\psi_2) } $. Since $\cf \times \Omega^1 (M ,S_1\oplus S_2 )$ is a real space we require the pulled back metric hereon to be real. For example, if we have an inner product which on $\Omega^1 (M,S_1\oplus S_2)$ is given by the fiberwise standard inner product on $S_1\oplus S_2$ and on the one forms combined with integrating over $M$ then this is fulfilled for any $(\psi_1,\psi_2)$  being an orthonormal basis for $\mathbb{C}^2$ in each point, since for $g_1,g_2\in \mathfrak{su}(2)$ we have 
\begin{eqnarray*}  ( g_1(\psi_1,\psi_2),g_2(\psi_1,\psi_2)    )   &=&  ( g_1\psi_1,g_2\psi_1    )+( g_1\psi_2,g_2\psi_2    )  \\
&= &( g_2^*g_1\psi_1,\psi_1    )+( g_2^*g_1\psi_2,\psi_2    )=Tr(g_2^*g_1) .  
\end{eqnarray*}
In particular, in this case the inner product is independent of the choice of $(\psi_1,\psi_2)$.  
In the case of a Sobolov inner product, i.e. an inner product of the form 
\begin{eqnarray*}
\langle   \xi_i  ,  \xi_j   \rangle_{p}  &=& \langle   \Psi_{(\psi_1,\psi_2 ) } (\xi_i),  \Psi_{(\psi_1,\psi_2) } (\xi_j)  \rangle_{p} \\
&=& \int_M ( (1+\Delta^p) (\xi_i \psi_1) ,(1+\Delta^p) (\xi_j \psi_1)) \\
&& +   ((1+\Delta^p) (\xi_i \psi_2) ,(1+\Delta^p) (\xi_j \psi_2))     ) dx ,
\end{eqnarray*}
which is the type of inner product that was used in the metric on $\cf$ constructed in \cite{Aastrup:2023jfw},
the product is not in general independent of the choice of $(\psi_1,\psi_2)$. If we for example take a different othonormal basis $(\psi_1',\psi_2')$ this amount to choosing a unitary matrix $N$ with $\psi_1'=N\psi_1$ and $\psi_2'=N\psi_2$. We thus get 
\begin{eqnarray*}
\langle   \xi_i  ,  \xi_j   \rangle_{p}  &=& \langle   \chi_{(\psi_1',\psi_2') } (\xi_i),  \chi_{(\psi_1',\psi_2') } (\xi_j)  \rangle_{p} \\
&=& \int_M ( (1+\Delta^p) (\xi_i N\psi_1) ,(1+\Delta^p) (\xi_j N\psi_1)) \\
&& +   ((1+\Delta^p) (\xi_i N\psi_2) ,(1+\Delta^p) (\xi_j N\psi_2))     ) dx \\
&=& \int_M (N^{-1} (1+\Delta^p) (\xi_i N\psi_1) ,N^{-1}(1+\Delta^p) (\xi_j N\psi_1)) \\
&& +   (N^{-1} (1+\Delta^p) (\xi_i N\psi_2) ,N^{-1} (1+\Delta^p) (\xi_j N\psi_2))     ) dx.
\end{eqnarray*}
In the case where we have a flat metric on $M$ it suffices that $N$ is constant in order for the inner product to be independent of the choice of $(\psi_1,\psi_2)$. For a non-trivial metric on $M$ the requirement would be that $N$ lies in the kernel of $\Delta$.

We refer the reader to \cite{Aastrup:2023jfw}  for details on how a metric that is compatible with the construction of a Dirac operator can be rigorously constructed on $\cf$.





\section{The Dirac operator}

The aim in this section is to construct a Dirac operator. To this end we first construct the CAR algebra.

\subsection{Constructing the CAR algebra}

We fix a metric $\langle \cdot,\cdot \rangle_{\nabla}$ in  (\ref{innerspinor}). Since 
$$\Psi_{\psi_1,\psi_2}  : \cf \times \Omega^1(M,\mathfrak{g}) \to\cf \times \Omega^1(M,S_1\oplus S_2) $$
is injective, we can consider  $\cf \times \Omega^1(M,\mathfrak{g})$ as a subspace of $\cf \times \Omega^1(M,S_1\oplus S_2)$ and hence it inherits the metric  from $\cf \times \Omega^1(M,S_1\oplus S_2)$. We choose an orthonormal basis $\{\xi_i \}$ of $\Omega^1(M,\mathfrak{g})$. Note that the $\xi_i$'s depends on $\nabla$. Next we extend this to an orthonormal basis $\{  \psi_i \}$ for  $\Omega^1(M,S_1\oplus S_2)$. 

We define  the CAR bundle over the configuration space of spinor-valued one-forms in $\OO^1(M,S_1\oplus  S_2)$ via the Fock space $\bigwedge^* \OO^1(M,S_1 \oplus S_2)$. 
Denote by $\mbox{ext}(\psi)$ the operator of external multiplication with $\psi\in \OO^1(M,S_1\oplus  S_2)$ on $\bigwedge^*\OO^1(M,S_1 \oplus S_2)$, and denote by $\mbox{int}(\psi)$ its adjoint, i.e. the interior multiplication by $\psi$: 
\begin{eqnarray}
\mbox{ext}(\psi) (\psi_1 \wedge  \ldots \wedge \psi_n) &=&  \psi\wedge \psi_1 \wedge  \ldots \wedge \psi_n,
\nn\\
\mbox{int} (\psi) (\psi_1 \wedge  \ldots \wedge \psi_n) &=& \sum_i (-1)^{i-1} \langle \psi, \psi_i \rangle_{\nabla} \psi_1 \wedge \ldots \wedge \psi_{i-1} \wedge \psi_{i+1} \ldots \wedge \psi_n,
\nn
\end{eqnarray}
where $\psi, \psi_i\in \OO^1(M,S_1\oplus  S_2)$. We have the following relations:
\begin{eqnarray}
\{\mbox{ext}(\psi_1), \mbox{ext}(\psi_2)  \} &=& 0,
\nn\\
\{\mbox{int}(\psi_1), \mbox{int}(\psi_2)  \} &=& 0,
\nn\\
\{\mbox{ext}(\psi_1), \mbox{int}(\psi_2)  \} &=& \langle \psi_1, \psi_2 \rangle_{\nabla}  
\nn%
\end{eqnarray}
as well as
$$
\mbox{ext}(\psi)^* = \mbox{int}(\psi),\quad \mbox{int}(\psi)^* = \mbox{ext}(\psi),
$$
where $\{\cdot,\cdot\}$ is the anti-commutator. We define  the Clifford multiplication operators $\bar{c}(\psi)$ and $c(\psi)$ given by
\begin{eqnarray}
  c(\psi) &=& \mbox{ext}(\psi) + \mbox{int}(\psi),
\nn\\
 \bar{c}(\psi) &=& \mbox{ext}(\psi) - \mbox{int}(\psi) 
\nn%
\end{eqnarray}
that satisfy the relations 
\begin{eqnarray}
 \{c(\psi_i), \bar{c}(\psi_j)\} &=& 0, \nn\\
 \{c(\psi_i), c(\psi_j)\} &=& \d_{ij}, \nn\\
 \{\bar{c}(\psi_i), \bar{c}(\psi_j)\} &=&- \d_{ij},
\nn
\end{eqnarray}
as well as
$$
c(\psi_i)^*= c(\psi_i), \quad \bar{c}(\psi_i)^* = - \bar{c}(\psi_i).
$$


Notice finally that since the inner product (\ref{innerspinor}) depends on $\nabla$ so does the basis $\{\psi_i\}$ and hence also the Clifford algebra. This means that the commutators between elements of the Clifford algebra and vectors $\frac{\pa}{\pa \xi_i}$ do not vanish\footnote{Strictly speaking we can here only derive in the directions $\xi_i$ which are in parallel to $\cf$. As already mentioned a discussion of this issue necessitates a BRST quantisation procedure adapted to our setup. We did this in \cite{Aastrup:2023jfw}. Throughout this paper we shall ignore this issue and refer the reader to \cite{Aastrup:2023jfw} for details.}
\begin{equation}
\left[ \frac{\pa}{\pa \xi_i}, z \right]\not = 0 ,\quad  z  \in \{c_j, \bar{c}_j,   \ldots \}. 
\label{nonzero} 
\end{equation}


\subsection{The Dirac operator}

We are now ready to construct a Dirac operator on the Hilbert space
$$
\ch_1 \oplus \ch_2= \left( L^2(\cf_1) \oplus L^2(\cf_2)  \right) \otimes \bigwedge^* \OO^1(M,S_1 \oplus S_2),
$$
where $\cf_1$ and $\cf_2$ are two copies of the space $\cf$.
We first define an action of the charge conjugation operator $C$ on $\bigwedge^* \OO^1(M,S_1 \oplus S_2)$
$$
C (\psi_1 \wedge \ldots \wedge \psi_n ) = C(\psi_1) \wedge \ldots \wedge C(\psi_n).
$$
Note that
$$
C^2 = (-1)^{\mbox{\tiny deg$(n)$}}, 
$$
where $n$ is the number of particles in the state on which $C$ acts and $\mbox{deg}(n)\in\{0,1\}$ is one on states with an odd particle number and zero on states with an even particle number. Futhermore we have 
\begin{equation}
C \bar{c}(\psi) C= (-1)^{\mbox{\tiny deg$(n)$}}\bar{c}(C(\psi))   .
\nn
\end{equation}
This follows from the relation 
$$ \langle \psi, \xi \rangle_{\nabla}=\overline{\langle C(\psi), C(\xi) \rangle}_{\nabla}.$$
We also have
$$
C^* C = C C^* =1,
$$
and since $C^* = (-1)^{\mbox{\tiny deg$(n)$}} C$ we get
\begin{equation}
C^* \bar{c}(\psi) C^*  =
-C \bar{c}(\psi) C  =
(-1)^{\mbox{\tiny deg$(n)+1$}}\bar{c}(C(\psi))  .
\end{equation}
On $\ch_1 \oplus \ch_2$ we define 
\begin{equation}
\cj =
\left(
\begin{array}{cc}
0 & C \\
C & 0
\end{array}
\right) \;\hbox{ with }\; \cj^2=(-1)^{\mbox{\tiny deg$(n)$}}.
\nn
\end{equation}
 %
We define the two Dirac operators on $\ch_1 $ and $ \ch_2$ respectively,
$$
D^+ = \sum_i \bar{c}(\psi_i) \nabla_{\xi_i}, \quad D^- = \sum \bar{c}(C(\psi_i)) \nabla_{\xi_i} ,
$$
where  $\nabla$ and denotes the levi-Civita connections associated with the metris on $\cf \times \Omega^1 (M,S_1\oplus S_2)$,
and then on $\ch_1 \oplus \ch_2 $
$$
\cd =
\left(
\begin{array}{cc}
D^+ & 0 \\
0 & D^-
\end{array}
\right).
$$
We find that 
\begin{equation}
\cj \cd \cj =  (-1)^{\mbox{\tiny deg$(n)$}} \gamma \cd
\label{operat}
\end{equation}
with
$$
\gamma =
\begin{pmatrix} -1 & 0 \\
0 & 1
\end{pmatrix} .
$$

\section{A rotation into Yang-Mills theory}

Let us now consider the kernel of $\cd$. Denote by $\eta^\pm$ elements in the kernel of $D^\pm$ and define
\begin{equation}
\Psi = 
\left(
\begin{array}{c}
\eta^+ \\
\eta^-
\end{array}
\right) \otimes \vert 0 \rangle
\label{regnvejr...}
\end{equation}
where $\vert 0 \rangle$ is the zero-particle state in $\bigwedge^* \OO^1(M,S_1 \oplus S_2)$. This gives us the equation
\begin{equation}
\cd \Psi =0
\label{Kamela}
\end{equation}
which can be interpreted as a Dirac equation on $\cf$. 
Next, let $U$ be a unitary operator acting in $\ch$ and consider the rotation of (\ref{Kamela})
$$
\cd^{\mbox{\tiny $U$}} \Psi^{\mbox{\tiny $U$}} =0, 
$$
where $\cd^{\mbox{\tiny $U$}} = U\cd U^*$ and $\Psi^{\mbox{\tiny $U$}} = U\Psi$. Specifically, consider the operator
$$
U =
\left(
\begin{array}{cc}
e^{i k CS(A)} & 0 \\
0 & e^{-i k CS(A)}
\end{array}
\right), 
$$
where $CS(A)$ is the Chern-Simons term
\begin{equation}
CS(A) =  \int_M \mbox{Tr} \left( {A}\wedge d{A} + \frac{2}{3} {A}\wedge {A} \wedge {A}\  \right)
\nn
\end{equation}
and where we let $k$ be an integer divided by $4\p$, which makes $U$ gauge invariant.
Next we write 
\begin{eqnarray}
\cd^{\mbox{\tiny $U$}} &=& \cd - [\cd, U] U^*
\nn\\
 &=& \left(
\begin{array}{cc}
D^+ - i k [D^+, CS(A)] & 0 \\
0 & D^- + i k [ D^-, CS(A)]
\end{array}
\right).
\nn
\end{eqnarray}
We are going to compute the square of $\cd^{\mbox{\tiny $U$}}$ and for simplicity we shall assume that $\nabla_{\xi_i}= \frac{\pa}{\pa\xi_i}$. We first write 
\begin{eqnarray}
\left(\cd^{\mbox{\tiny $U$}}\right)^2 &=&  
\left(
\begin{array}{cc}
e^{ikCS} \left(D^+\right)^2 e^{-ikCS}  & 0 \\
0 & e^{-ikCS} \left(D^-\right)^2 e^{ikCS} 
\end{array}
\right) + \X
\label{mrichter}
\end{eqnarray}
where $\X$ is an additional term due to (\ref{nonzero}), and then compute
\begin{eqnarray}
    e^{\pm ikCS} \left(D^\pm\right)^2 e^{\mp ikCS} 
&=& k^2\sum_{i} \bigg(\pm \frac{i}{k} \frac{\pa^2 CS(A)}{\pa \xi_i \pa \xi_i} 
+ \left( \frac{\pa CS(A)}{\pa \xi_i}\right)^2 
\nn\\
&&\pm \frac{2i}{k} \frac{\pa CS(A)}{\pa \xi_i} \frac{\pa}{\pa \xi_i}
-  \frac{1}{k^2}\left(\frac{\pa}{\pa \xi_i}\right)^2 \bigg).  
\nn
\end{eqnarray}
We then use
\begin{equation}
\frac{\pa CS}{\pa \xi_i} =  2 \int_M \mbox{Tr}\left(\xi_i\wedge F(A)  \right)
\nn
\end{equation}
where $F(A)$ is the field strength tensor of the connection $A$, as well as the definition of a field operator (see \cite{Aastrup:2020jcf})
\begin{equation}
\hat{E}_i=\frac{i}{2k}\frac{\partial }{\partial \xi_j},\quad
\hat{E}_A({\bf m})=\frac{i}{2k}\sum_j \xi_j({\bf m}) \hat{E}_i ,
\nn
\end{equation}
where the index '$A$' indicates that the vectors $\xi_i$ generally depend on the connection $A$, to obtain 
\begin{eqnarray}
e^{\pm ikCS} \left(D^\pm\right)^2 e^{\mp ikCS} &=&
 4k^2\Bigg(  \left( \hat{E}_i \right)^2 +  \left(\int_M \mbox{Tr}\left(\xi_i\wedge F(A) \right)\right)^2 
\nn\\&&
\pm   2  \int_M \mbox{Tr}\left( F(A) \wedge\hat{E}_A  \right)
\pm \mbox{Tr}_\xi \left(i\nabla^A \right)\Bigg) .
\nn
\end{eqnarray}
Here
\begin{equation}
\mbox{Tr}_\xi \left(i\nabla^A \right) = i\sum_i \mbox{Tr}\left(\xi_i  \nabla^A \xi_i \right)
\label{heron}
\end{equation}
is a spectral invariant, which we first discussed in section 5.3 in \cite{Aastrup:2020jcf} and which is related to the eta-invariant for $\nabla^A$ that was first introduced by Atiyah, Patodi, and Singer \cite{Atiyah}-\cite{AtiyahIII}. This spectral invariant measures the assymmetry of the spectrum of $\nabla^A$. 
In total we therefore find 
\begin{eqnarray}
\left(\cd^{\mbox{\tiny $U$}}\right)^2 =  
\left(
\begin{array}{cc}
H^{(+)}_{\mbox{\tiny YM}} + \mbox{Tr}_\xi \left(i\nabla^A \right)  & 0 \\
0 & H^{(-)}_{\mbox{\tiny YM}}  - \mbox{Tr}_\xi \left(i\nabla^A \right)
\end{array}
\right) + \mbox{curvature terms} + \X
\nn
\end{eqnarray}
where
\begin{eqnarray}
    H^{(\pm)}_{\mbox{\tiny YM}} = 
    4k^2\Bigg(  \left( \hat{E}_i \right)^2 +  \left(\int_M \mbox{Tr}\left(\xi_i\wedge F(A) \right)\right)^2 
\pm   2  \int_M \mbox{Tr}\left( F(A) \wedge\hat{E}_A  \right)\Bigg)
\nn
\label{whatever2}
\end{eqnarray}
Let us compare $H^{(\pm)}_{\mbox{\tiny YM}}$ to a Langrangian setup in a local limit. If we write the Yang-Mills action 
\begin{eqnarray}
S_{\mbox{\tiny YM}}
= S^{(+)}_{\mbox{\tiny YM}} + S^{(-)}_{\mbox{\tiny YM}} 
\nn
\end{eqnarray}
with
$$
S^{(\pm)}_{\mbox{\tiny YM}}=\frac{1}{2}\int_\cm \mbox{Tr}\left( {\bf F}\wedge \left( \star {\bf F} \pm \theta {\bf F}\right)\right)
$$
where $\cm$ is now a four-dimensional manifold in which $M$ is a Cauchy surface, ${\bf F}$ is a four-dimensional field-strength tensor, and $\star$ is the four-dimensional Hodge dual, then $H^{(\pm)}_{\mbox{\tiny YM}}$ corresponds to the selfdual and anti-selfdual sectors of an $SU(2)$ Yang-Mills theory in the sense that
$$
H_{\mbox{\tiny YM}} = H^{(+)}_{\mbox{\tiny YM}}+H^{(-)}_{\mbox{\tiny YM}}.
$$

To make this point clearer we can write the $F^2$-term in (\ref{whatever2}) in a local limit, where the integral kernel 
$$
K({\bf m}_1,{\bf m}_2)=\sum_i \xi_i({\bf m}_1) \xi_i ({\bf m}_2)
$$
gives us a Dirac delta function (for details on the connection to Yang-Mills quantum field theory we refer the reader to \cite{Aastrup:2020jcf}), as
$$
 \left(\int_M \mbox{Tr}\left(\xi_i\wedge F(A) \right)\right)^2 \Bigg\vert_{\mbox {\tiny local limit}   }= \int_M \mbox{Tr}\left( F(A)^2 \right).
$$
Also, if we introduce the local field operator
$$
\hat{A}({\bf m})= \sum_i x_i \xi_i({\bf m})
$$
then we obtain the commutator relation
$$
[\hat{E}_A({\bf m_1}),\hat{A}({\bf m_2})]
=K({\bf m}_1,{\bf m}_2),
$$
which in a local limit gives us the canonical commutation relations of a $SU(2)$ quantum gauge theory. 
Adding all this up we conclude that we obtain a candidate for a non-perturbative quantum Yang-Mills theory.\\

Before we end this section let us point out that the ground state (\ref{regnvejr...}) is degenerate. If, for instance, we consider the trivial geometry on $\cf$ where $\nabla_{\xi_i}= \frac{\pa}{\pa \xi_i}$, then the state
\begin{equation}
\Psi' = 
\left(
\begin{array}{c}
\eta^+ \\
\eta^-
\end{array}
\right) \otimes \Psi_{\mbox{\tiny CAR}}
\nn
\end{equation}
where $\Psi_{\mbox{\tiny CAR}}$ is an arbitrary element in the Fock space $\bigwedge^*\OO^1(M,S_1 \oplus S_2)$, will also lie in the kernel of $\cd$. In other words, in this particular case the ground state will be infinitely degenerate. When the geometry is not trival the situation is more complicated since $\cd$ will also interact with the Fock space, but we still suspect a certain level of degeneracy will remain.




\section{Discussion}

In this paper we have shown that given a Dirac operator on a configuration space of gauge connections one can obtain the key building blocks of a Yang-Mills quantum field theory via a unitary transformation that involves a Chern-Simons term. This basic result suggests that the longstanding question about how to rigorously construct quantum Yang-Mills theory non-perturbatively might be successfully reformulated as a question of rigorously defining a Dirac operator on a configuration space. We have previously discussed this question at length (see \cite{Aastrup:2023jfw} for the most recent update) and found that it essentially boils down to two key issues: convergence and the Gribov ambiguity \cite{Gribov:1977wm}. Concerning convergence, then we have shown in \cite{Aastrup:2023jfw} that it is possible in certain cases to rigorously construct a metric on a configuration space that is compatible with the construction of a Dirac operator. A central ingredient in such a construction is a type of Sobolev-norm that regulates the ultra-violet limit and essentially translates into a choice between unitarily non-equivalent representations of the $\mathbf{HD}$-algebra. The main obstruction to a widening of this result is the Gribov ambiguity. In \cite{Aastrup:2023jfw} we did, however, propose a novel approach to the resolution of this obstruction.

Strictly speaking one does not need a Dirac operator on the configuration space to get to Yang-Mills theory; a Laplace operator would suffice (see equation (\ref{mrichter}). There are several reasons why we emphasise the Dirac operator: first of all, the spectral invariant $\mbox{Tr}_\xi \left(i\nabla^A \right)$ in equation (\ref{heron}) comes from a term that involves two derivations of the Chern-Simons term. In the case where one works with a Bott-Dirac operator instead of a Dirac operator it is terms of this type that gives rise to a fermionic sector, a feature that would be lost if one were to use a Laplace operator. Also, as already mentioned, the machinery of noncommutative geometry is one of unification, both in terms of a bosonic sector, as has been demonstrated in the case of the standard model, where the entire bosonic sector emerges from an inner fluctuation of the Dirac operator used in the spectral triple formulation of the model \cite{Connes:1996gi}, and in terms of a unification between bosons and fermions, as is the case with the aforementioned Bott-Dirac operator on a configuration space. All of this structure would be lost if one were to use a Laplace operator.

Concerning the inclusion of half-integer spin fermions then it is interesting whether the mapping of $Cl(3)$ into $Cl(4)$ in section 2.2 corresponds to a choice of space-time foliation and thus a choice of lapse and shift fields. Indeed, it seems likely that it is possible to choice different embeddings at different scales, which could be interpreted in terms of a non-trivial space-time foliation.

An important question is to what extend our construction depends on a Rimannian metric on the underlying manifold. This is related to the question as to what role general relativity plays in this setup. We have previously suggested that a metric on the underlying manifold might be emergent since it is encoded into a metric on the configuration space. In other words, the idea is to consider dynamical metric on the configuration space and then study semi-classical states from which a metric on the manifold might emerge (see \cite{Aastrup:2023jfw} for a more detailed discussion). Concerning the present construction it is clear that the spin-bundle does involve metric information and hence the construction of the Dirac operator carried out in this paper does involve a metric on the underlying manifold. It is conceivable, however, that one can construct the Dirac operator without this background information: instead of the spin-bundle one can just use two copies of $\mathbb{C}$.


Finally, having constructed a real structure it is an interesting question what 
KO-dimension our construction might have (see page 11 of \cite{Connes:1996gi}). We find, however, that our construction does not match the setup described in \cite{Connes:1996gi}; specifically, equation (\ref{operat}) does not show that the Dirac operator either commutes or anti-commutes with the real structure but rather that their interaction depends on the number of particles in the state on which the they act. Also, the presence of the matrix $\gamma$ suggest that we could have a construction that consist of two spaces of different dimensionality. This situation should, perhaps, not be surprising since our construction involves an infinite-dimensional configuration space and a spectral triple-like construction that has clear ties to both bosonic and fermionic quantum field theory and hence, potentially, to space and time involving a Minkowski signature.

\vspace{1cm}
\noindent{\bf\large Acknowledgements}\\

\noindent

JMG would like to express his gratitude to entrepreneur Kasper Bloch Gevaldig for his unwavering financial support. JMG is also indepted to Master of Science in Engineering Vladimir Zakharov and to the following list of sponsors for their generous support:   Frank Jumppanen Andersen,
Bart De Boeck, Simon Chislett,
Jos van Egmond,
Trevor Elkington,
Jos Gubbels,
Claus Hansen,
David Hershberger,
Ilyas Khan,
Simon Kitson,
Hans-J\o rgen Mogensen,
Stephan M{\"u}hlstrasser,
Bert Petersen,
Ben Tesch,
Jeppe Trautner,
and the company
Providential Stuff LLC. JMG would also like to express his gratitude to the Institute of Analysis at the Gottfried Wilhelm Leibniz University in Hannover, Germany, for kind hospitality during numerous visits.

\end{document}